\documentclass[12pt, letterpaper]{article}
\usepackage[utf8]{inputenc}
\usepackage[T1]{fontenc}
\usepackage{amsmath}
\usepackage{amsfonts}
\usepackage{biblatex}
\usepackage{mathrsfs}
\usepackage{mathtools}
\usepackage{tikz}
\usetikzlibrary{matrix}
\numberwithin{equation}{section}
\usepackage{multirow}
\usepackage{hyperref}

\usepackage{varioref}
\usepackage[noabbrev]{cleveref}
\usepackage{float}
\usepackage{lmodern}
\usepackage{palatino}
\usepackage{mathptmx}
\usepackage[scaled=.90]{helvet}
\usepackage[affil-it]{authblk}
\usepackage{xcolor}

\begin{document}
\title{\textrm{  D-brane Superpotentials and Geometric Invariants in Complete Intersection Calabi-Yau Manifolds }}
\date{}
\author[]{Xuan Li \footnote{E-mail:\href{mailto: lixuan191@mails.ucas.ac.cn}{ lixuan191@mails.ucas.ac.cn}}, Yuan-Chun Jing and Fu-Zhong Yang}
\affil[]{School of Physical Sciences,University of Chinese Academy of Sciences,\\No.19(A) Yuquan Road, Shijingshan District, Beijing, P.R.China 100049}
\maketitle

\begin{abstract}
\textbf{Abstract}: 
 By blowing up the ambient space along the curve wrapped by B-branes, we study the brane superpotentials and Ooguri-Vafa invariants on complete intersections Calabi-Yau threefolds. On the topological  B-model side,  B-brane superpotentials  are expressed in terms of  the period integral of the blow-up manifolds. By mirror maps, the superpotentials are generating functions of  Ooguri-Vafa invariants counting holomorphic disks on the topological  A-model side.    
\end{abstract}

\newpage
\tableofcontents
\newpage

\section{Introduction}  
 For open topological string theory,  A-models admit A-branes wrapping special Lagrangian cycles on $M^*$ described by Fukaya category, and B-models admit B-branes wrapping holomorphic cycles on $M$ described by derived category of coherent sheaves\cite{Kontsevich1994}.  Then, open mirror symmetry implies the category equivalence $\mathrm{Fuk}(M^*)=\mathrm{D^b}(M)$.
 
 As  the topological sector of the superpymmetric spacetime effective Lagrangian, D-brane superpotential plays an important role in describing string theory compactifications.  It can be obtained by dimension reduction of  the holomorphic Chern-Simons theory  \cite{Witten1992} or analysis of $A_\infty$-structure\cite{Aspinwall2005,Aspinwall2004} in derived category.
 The  A-brane superpotential is generated by disk instantons\cite{Kachru2000,Kachru1999} and its expansion at the large volume point encodes   the Ooguri-Vafa invariants counting the holomorphic disks ending on A-branes\cite{Ooguri1999}.  The B-brane superpotential arises from variation of mixed Hodge structure, which depends on the brane moduli  and the complex structure  moduli of the Calabi-Yau threefold. It satisfied the extended Picard-Fuchs differential system \cite{Alim2009,Forbes2005,Li2009} and measures the obstructions to deforming D-brane  in the open string direction  \cite{Herbst2004}. 
 
 Superpotentials and  Ooguri-Vafa invariant for toric branes  on compact Calabi-Yau threefolds $M$ in \cite{Alim2009}. The B-brane geometry is captured by an auxiliary divisor $D$. The  extended Picard-Fuchs equation associated to a enhanced polyhedron, arises from
 N=1 special geometry\cite{Forbes2003,Lerche2002}
, and can be  obtained by Griffish-Dwork reduction method \cite{morrison1993,Schmid1973} or GKZ-system\cite{Hosono1995,Gelfand1990}, governing the variation of Hodge structure of $(M,D)$\cite{Jockers2008}. It is argued in  \cite{Grimm2010,Grimm2008} that  
under blowing up calabi-Yau threefold $M$ along curve $\Sigma$ wrapped by B-branes,  the deformation theory of the blow-up manifold $M^\prime$  with an exceptional divisor $E$  is equivalent to the deformation theory of the original manifold $M$ with the curve $\Sigma$.Then,  the B-brane superpotential can be expressed in terms of period integral of  $X^\prime$.  In this work, we mainly study the B-brane superpotentials  and Ooguri-Vafa invariants on compact complete intersection Calabi-Yaus (CICY) threefolds  by blow-up constructions.  Periods integrals are solved by GKZ-system, and then used to find mirror maps and superpotentials.  Also, Ooguri-Vafa invariants are extracted from the A-brane superpotentials at large volume point .

  The organization of this paper is as follows. In section 2, we review the construction  of complete intersection Calabi-Yau  manfolds in  weighted projective space and variation of mixed Hodge structure, and outline the procedure to blow up a curve on a Calabi-Yau threefold.  In section 3 , for one-parameter CICYs($M_{2,4}$,$M_{3,4}$), we apply the blowing-up method and obtain the new manifolds. The Picard-Fuchs equations and their solutions are derived by GKZ hypergeometric system from toric information of the enhanced polyhedron. The superpotential are identified as double-logarithmic solutions of the Picard-Fuchs equations and Ooguri-Vafa invariants are extracted at large volume phase of A-model. The last section is a brief summary and outlook for further research. In Appendix A, we summarize the GKZ-system of blow-ups resulting from blowing up  curves on $M_{2,6},M_{3,6},M_{4,6}$. In Appendix B, we present the Ooguri-Vafa invariants on aforementioned  models.

\section{Toric Geometry and Blowing up}
\subsection{Complete Intersection Calabi-Yau and GKZ-System}
The complete intersection Calabi-Yau threefolds  $M^*\subset\mathbb{P}^5_{(\omega_1,\omega_2,\omega_3,\omega_4,\omega_5)}$ considered here are described by the reflexive polyhedron $\Delta=\Delta_1\times \Delta_2$\cite{Batyrev1994,borisov1993}. The mirror threefold $M_{d_1,d_2}$ is the intersection of the vanishing locus of two homogeneous polynomials of degree $d1$ and $d_2$, captured by  the combinatorial data of the dual polyhedron $\Delta^*$ with vertices $\nu^*_{1}=(1,0,\ldots,0),\ldots,\nu^*_{5}=(0,\ldots,0,1),\nu^*_6=(-\omega_1,-\omega_2,-\omega_3,-\omega_4,-\omega_5)$.

The vertices are partitate into two group,  $\{\nu^*_{i}\}=E_1  \cup E_2$, where $E_i$ contains $d_i$  vertices, known as the nef-partition.    Each vertex $\nu^*_{i}$ of $E_i$ is extended to $\bar{\nu}^*_{i}=(e^{(i)},\nu^*_i)$ . With  adding extra vertices $\bar{\nu}^*_{0,i}=(e^{(i)},0)$, the new vertices $\bar{\nu}^*_{i}$ satisfy the relations,
\begin{equation*}
\sum l^i\bar{\nu}^*_i,=0,\quad l=(-d_1,-d_2;\omega_1,\omega_2,\omega_3,\omega_4,\omega_5),
\end{equation*}

The mirror manifold $M_{d_1,d_2}$ is defined by the following Laurent polynomials in the torus coordinates $X_i$\cite{Batyrev1995,Hosono1994},
\begin{equation}\label{eq:2.1}
P_r=a_{0,r}-\sum_{\nu^*_{i}\in E_r}a_{i}X^{\nu^*_{i}}, r=1,2
\end{equation}
The period integral of $M$,
\begin{equation*}
    \varpi(a)=\int\frac{a_1\cdots a_6}{P_1 P_2}\prod_{i=1}^5\frac{dX_i}{X_i},
\end{equation*}
is annihilated by the following GKZ-system,
\begin{equation}\label{eq:2.2}
\begin{gathered}
\mathcal{L}=\prod_{l_j>0}(\frac{\partial}{\partial a_j})^{l_j}-\prod_{l_j<0}(\frac{\partial}{\partial a_j})^{-l_j},\\[10pt]
\mathcal{Z}_i=\sum_j(\bar{v}_j)^i \vartheta_j-\beta_i, \quad i=0,\ldots,5
\end{gathered}
\end{equation}
Here $\beta$ is the  exponent, $\vartheta_j=a_j \frac{\partial}{\partial a_j}$  is the logarithmic derivative and $l_j$ corresponds to the maximal triangulation of $\Delta^*$.

For open topological string theory, the A-brane configuration on $M^*$ is specified by $l$ with another two relation $l^i,i=1,2$,such that $\sum_j l^i_j=0$\cite{Aganagic2001,Aganagic2000}. The corresponding B-brane geometry is described by the curve $\Sigma$ on the mirror threefold $M$,
\begin{equation*}
    \Sigma:\{P_1=P_2=0\}\cap \{Q(D_1)=0\}\cap\{Q(D_2)=0\},
\end{equation*}
where $Q(D)$ are divisors defined by vertices on polyhedron $\Delta^*$. The polyhedron $\Delta$ can be extended to a higher dimensional polyhedron $\Delta^\prime$, called the enhanced polyhedron, by adding vertices on the original vertices $\Delta$. 
The GKZ- system associated to the enhanced polyhedron annihilates the relative preiod integral and  is used as a way to understand brane geometry in \cite{Alim2009}. 

The coordinates on open-closed moduli space is given by 
\begin{equation}\label{eq:2.3}
z_j=(-a_0)^{l^j_0} \prod_i a_i^{l^j_0},
\end{equation}
For appropriate choice of basis vector $l^j$, solutions to the GKZ system can be written in terms of the deformed gamma series,
\begin{equation*}
\varpi(z;\rho)=\sum \frac{\Gamma(1-\sum_j l^j_0(n_j+\rho_j))}{\prod_{i>0}\Gamma(1+\sum_j l^j_i(n_j+\rho_j))}\prod_k z_k^{n_j+\rho_j}
\end{equation*}
 then we have a natural basis for the period integral,
 \begin{equation}\label{eq:2.6}
 \begin{aligned}
 \omega_0(z)&=\varpi(z;\rho)|_{\rho \rightarrow 0},\\[10pt]
 \omega_{1,i}(z)&=\partial_{\rho_i}\varpi(z;\rho)|_{\rho \rightarrow 0},\\[10pt]\omega_{2,i}(z)&=\sum_{j,k}K_{ijk}\partial_{\rho_j}\partial_{\rho_k}\varpi(z;\rho)|_{\rho \rightarrow 0}\\
 ...
 \end{aligned}
 \end{equation}
 and at the large complex structure point $z = 0$, 
  \begin{equation*}
 \begin{aligned}
 \omega_0(z)&\sim1 + \mathcal{O}(z),\\[10pt]
 \omega_{1,i}(z)& \sim \log(z_i),\\[10pt]\omega_{2,i}(z)& \sim \log(z_j)\log(z_k),\\
 \end{aligned}
 \end{equation*}
 Here $\omega_{1,i}$ define the open-closed mirror map by
\begin{equation}\label{eq:2.5}
t_i(z)=\frac{\omega_{1,i}(z)}{\omega_0(z)},\quad q_i=e^{2 \pi i t_i},
\end{equation}
and  the special solution $\Pi=\mathcal{W}_{open}(z)$ has  instanton expansion near the large volume point, containing the Ooguri-Vafa invariants of the A-brane geometry,
\begin{equation*}
\mathcal{W}_{inst}(q)=\sum_\beta \sum_{k=1}^\infty N_\beta \frac{q^{k \cdot\beta}}{k^2}.
\end{equation*}

\subsection{Relative Period and Variation of Mixed Hodge Structure}
For the submanifold $\Sigma$,  embedded by the map $i:\Sigma \hookrightarrow M$ in the Calabi-Yau threefold $M$, the space of relative forms $\Omega^*(M,\Sigma)$ is the subspace of forms $ \Omega(M)$, defined as the kernel of the pullback, $i^* : \Omega(M) \rightarrow  \Omega^*(\Sigma)$.  Consider the exact sequence,
\begin{equation*}
0 \rightarrow  \Omega^*(M,\Sigma) \overset{i^*}{\hookrightarrow} \Omega^*(M)
\rightarrow \Omega^*(\Sigma)\rightarrow 0 
\end{equation*}
Then, the relative cohomology groups $H^* (X,S)$  arise from the space of closed modulo
exact relative forms with respect to the de Rham differential and   the long
exact sequence for the three-form cohomology group is obtained,
\begin{equation*}
H^3(M,\Sigma) \cong ker(H^3(M)\rightarrow H^3(\Sigma)) \oplus coker(H^2(M)\rightarrow H^2(\Sigma)),
\end{equation*}
Thus, we can represent a relative three form $\underline{\Theta}$ as a pair of closed three form $\Theta$ and a closed two form $\theta$,
\begin{equation*}
\underline{\Theta}=(\Theta,\theta)\in H^3(M,\Sigma),
\end{equation*}
and obey the equivalence relation,
\begin{equation*}
\underline{\Theta} \sim \underline{\Theta}+(d \alpha,i^*\alpha-d \beta), 
\end{equation*}
where $\alpha$ is a two form on $M$, and $\beta$ is a one form on $\Sigma$. Similarly, the relative homology group $H_3(M,\Sigma)$ consists of relative three cycles $\underline{\Gamma}$ with boundaries $\partial \underline{\Gamma}$ lying in $S$. The the duality pairing between $\underline{\Gamma}$ and $\underline{\Theta}$ is given by,
\begin{equation*}
\int_{\underline{\Gamma}}\underline{\Theta} \equiv \int_{\underline{\Gamma}} \underline{\Theta}-\int_{\partial \underline{\Gamma}}\theta,
\end{equation*}
In particular, the relative period,
\begin{equation*}
\Pi^a(z,\hat{z})=\int_{\underline{\Gamma}^a} \underline{\Omega},
\end{equation*} 
where $\underline{\Omega}$ is a relative three form, and $\underline{\Gamma}^a$ is a relative three cycle. The moduli dependence of the period is captured by the variation of the mixed Hodge structure given by cohomology group $H^3(X,S,\mathbb{Z})$ and the filtration,
\begin{equation*}
\begin{aligned}
&F^3=H^{3,0}(M,\Sigma),\\[10pt]
&F^2=H^{3,0}(M,\Sigma) \oplus H^{2,1}(M,\Sigma),\\[10pt]
&F^1=H^{3,0}(M,\Sigma) \oplus H^{2,1}(X,\Sigma) \oplus H^{1,2}(M,\Sigma),\\[10pt]
&F^0=H^{3,0}(M,\Sigma) \oplus H^{2,1}(M,\Sigma) \oplus H^{1,2}(M,\Sigma) \oplus H^{0,3}(M,\Sigma).
\end{aligned}
\end{equation*} 
and the weight filtration induced from is
\begin{equation*}
W_3 \cong H^3(M),\quad W_4\cong H^3(M) \oplus H^2_{var}(\Sigma) \cong H^3(M,\Sigma),
\end{equation*} 
Notice that an infinitesimal closed -string complex structure deformation $\partial_z$ changes the Hodge type of a $(p,q)$-form and an infinitesimal open-string deformation $\partial_u$ affects the two-form sector $H^2_{var}(X)$, thus the two deformations as tangent vectors in the opn/closed  string moduli  space, act on the mixed Hodge structure as:  
\begin{equation*}
\begin{tikzpicture}
\matrix(m)[matrix of math nodes, row sep=0.5em,column sep=4em,minimum width=2em]
{
  F^3 \cap W_3 & F^2 \cap W_3 & F^1 \cap W^3 & F^0 \cap W^3  \\
  \quad  \\
  \quad& F^2 \cap W_4 & F^1 \cap W_4 &F^0 \cap W_4 \\
  };
  \path[-stealth]
  (m-1-1)  edge  node [above] {$\partial_z$}(m-1-2)
  (m-1-2)  edge  node [above] {$\partial_z$}(m-1-3)
  (m-1-3)  edge  node [above] {$\partial_z$}(m-1-4)
  (m-3-2)  edge  node [above] {$\partial_z,\partial_{\hat{z}}$}(m-3-3)
  (m-3-3)  edge  node [above] {$\partial_z,\partial_{\hat{z}}$}(m-3-4)
  (m-1-1)  edge  node [above] {$\partial_{\hat{z}}$}(m-3-2)
  (m-1-2)  edge  node [above] {$\partial_{\hat{z}}$}(m-3-3)
  (m-1-3)  edge  node [above] {$\partial_{\hat{z}}$}(m-3-4)
  (m-1-4)  edge  node [right] {$\partial_{\hat{z}}$}(m-3-4);
\end{tikzpicture}
\end{equation*}
and  the two form sector $H^2_{var}(X) \cong W_4/W_3$ constitutes a subsystem,
\begin{equation*}
F^2 \cap(W_4/W_3) \overset{\partial_z,\partial_{\hat{z}}}{\rightarrow} F^1 \cap(W_4/W_3)\overset{\partial_z,\partial_{\hat{z}}}{\rightarrow} F^0 \cap(W_4/W_3)
\end{equation*}
which is essential to derive and solve the Picard-Fuchs equations related to the relative period.

\subsection{Blowing Up}
The  construction and properties  of blowing up  a manifold along its submanifold is an integral part of this work. For more detail about blowing up, we refer to \cite{Hartshorne2013}.

The blow-up threefold $M^\prime$ resulting from blowing up the threefold $M$ along its submanifold $\Sigma$ is obtained by gluing local blow-ups.  Consider  multidisks  $U_\alpha$ on  $M$ with holomorphic coordinates $x_{\alpha,i},i=1,2,3$, and $V_\alpha$  specified by $x_{\alpha,1}=x_{\alpha,2}=0$ on each $U_\alpha$. The local blow-up is defined by,
\begin{equation*}
\tilde{U}_\alpha=\{(x_{\alpha,1},x_{\alpha,2},x_{\alpha,3},(y_1:y_2))\subset U_\alpha\times \mathbb{P}^1:x_{2,\alpha}y_1-x_{\alpha,1}y_2=0\},
\end{equation*}
where $y_1,y_2$ are the homogeneous coordinates on $\mathbb{P}^1$.  The manifold $\tilde{U}_\alpha$ is the blow-up of $U_\alpha$ along $V_\alpha$, under  the projection map $\pi_\alpha:\tilde{U}_\alpha \rightarrow U_\alpha$, with  the inverse image $E_\alpha=\pi_\alpha^{-1}(V_\alpha)$  an exceptional divisor of  $\tilde{U}_\alpha$.  The coordinates patches $U_1=\{y_1 \neq 0\}$ and  $U_2=\{y_2 \neq 0\}$ have holomorphic coordinates
\begin{equation*}
\begin{gathered}
z^{(1)}_1=x_{\alpha,1},\quad z_2^{(1)}=\frac{y_2}{y_1}=\frac{x_{\alpha,2}}{x_{\alpha,}},\quad y_3=x_{\alpha,3},\\[10pt]
z^{(2)}_1=\frac{y_1}{y_2}=\frac{x_{\alpha,1}}{x_{\alpha,2}},\quad z_2^{(1)}=x_{\alpha,2},\quad y_3=x_{\alpha,3},\\
\end{gathered}
\end{equation*} 
with transition function on $U_1 \cap U_2$,
\begin{equation*}
g_{ij}=z^{(j)}_i=\frac{y_i}{y_j}=\frac{x_{\alpha,i}}{x_{\alpha,j}},
\end{equation*}

 Let $\{U_\alpha\}$ be a collection of disks in $M$ covering $\Sigma$ such that $V_\alpha=\Sigma\cap U_\alpha$.
The local blow-ups $\tilde{U}_\alpha $ can be patched up to a manifold $\tilde{U}=\cup_{\pi_{\alpha \beta}}\tilde{U}_\alpha$ by the projection map,
\begin{equation*}
\pi_{\alpha \beta}:\pi_{\alpha \beta}^{-1}(U_\alpha \cap U_\beta)\rightarrow \pi_\beta^{-1}(U_\alpha \cap U_\beta),
\end{equation*}
$M^\prime=\tilde{U}\cup_\pi M-\Sigma$, together with  $\pi:X^\prime \rightarrow X$, is called the blow-up of $M$ along $\Sigma$, with exceptional divisor $E$.

 The excision theorem of cohomology \cite{Vick2012} implies that,
\begin{equation*}
H^3(M,\Sigma)\cong H^3(M-\Sigma)\cong H^3(M^\prime-E) \cong H^3(M^\prime,E)
\end{equation*}
which leads to the equivalence between  the mixed Hodge structures of $H^3(M,\Sigma)$ and the mixed Hodge structure $H^3(M^\prime,E)$ over the corresponding  moduli space. Thus, the blow-up $M^\prime$ can be used to compute  superpotentials with brane geometry $(M,\Sigma)$.

Given $M$ and $\Sigma$ as follows,
\begin{equation*}
\begin{aligned}
   &M:P_1=P=2=0,\\[10pt] &\Sigma:P_1=P_2=0,h_1=h_2=0,
   \end{aligned}
\end{equation*}
 where $h_1$ and $h_2$ are two divisors.   The blow-up manifold $M^\prime$ is given globally as the complete intersection in the total space of the projective bundle $\mathcal{W}=\mathbb{P}(\mathcal{O}(h_1)\oplus\mathcal{O}(h_2))$,
\begin{equation}
P_1=P_2=0, \quad Q \equiv y_1 h_2-  y_2 h_1=0
\end{equation}
where $(y_1,y_2)\sim\lambda(y_1,y_2)$ is the projective coordinates on the $\mathbb{P}^1$ -fiber of the blow-up $X^\prime$. 

The holomorphic three form $\Omega^\prime$ on $M^\prime$ is pullback of the holomorphic three form on $M$ by the projection map,
\begin{equation*}
    \Omega^\prime=\int_{S^1_P}\int_{S^1_Q}\frac{h_i}{y_i}\frac{\omega_{\mathbb{P}_1}}{Q}\wedge \frac{\omega}{P_1P_2},\quad i=1,2
\end{equation*}
Here $\omega$ and $\omega_{\mathbb{P}_1}$ denote the invarant holomorphic top form on the $M$ and $\mathbb{P}_1$,respectively. $S^1_P$ and $S^1_Q$ are small loops around $P_1=P_2=Q=0$ encircling only the corresponding poles. 

The period integral of $\Omega^\prime$ is annihilated by the GKZ-system associated with the enhanced polyhedron. This GKZ-system and explicit construction of the enhanced polyhedron will be present by our examples in details.

\section{Superpotentials and Ooguri-Vafa Invariants on one parameter CICY}
In this section, we study  one-parameter compact CICYs with curves wrapped by B-branes, and obtain B-brane superpotentials and Ooguri-Vafa invariants by GKZ-system.  
 \subsection{\texorpdfstring{Branes on CICY $M_{2,4}$}{M(2,4)} }
\subsubsection{GKZ-system and Blowing Up Geometry}
The  threefold $M^*$ associated to A-model is described by  the polyhedron $\Delta^*$ with one internal point $\nu^*=(0,0,0,0,0)$ and six vertices,
\begin{equation*}
\begin{gathered}
\nu^*_1=(1,0,0,0,0),
\quad\nu^*_2=(0,1,0,0,0),
\quad \nu^*_3=(0,0,1,0,0),\\[10pt]
\nu^*_4=(0,0,0,1,0),\quad\nu^*_5=(0,0,0,0,1),\quad \nu^*_6=(-1,-1,-1,-1,-1),
\end{gathered}
\end{equation*}
We group the vertices into two sets $E_1=\{\nu^*_1,\nu^*_2\}$ and $E_2=\{\nu^*_3,\nu^*_4,\nu^*_5,\nu^*_6\}$, extend the vertices $\bar{\nu}^*=(\overrightarrow{e}_{1,2};\nu^*)$ and add two vertices $\nu^*_{0,k}=(\overrightarrow{e}^{(k)},\overrightarrow{0})$. This lead to the defining equation of mirror Calabi-Yau $M_{2,4}$ as the following two Laurent polynomials by equation  \ref{eq:2.1},
\begin{equation*}
\begin{cases}
P_1=a_{0,1}-a_1 X_1-a_2X_2\\[10pt]
P_2=a_{0,2}-a_2 X_2-a_3X_3-a_4 X_4-a_5 X_5-a_6(X_1X_2X_3X_4X_5)^{-1}, 
\end{cases}
\end{equation*}
or in homogeneous coordinates \cite{Morrison1997},
\begin{equation*}
\begin{cases}
P_1=a_{0,1}x_3x_4x_5x_6-a_1 x_1^4-a_2 x_2^4\\[10pt]
P_2=a_{0,2}x_1 x_2 -a_3 x_3^2 -a_4 x_4^2-a_5x_5^2-a_6 x_6^2, 
\end{cases}
\end{equation*}
where $a$ is complex-valued free parameter. The
  linear relation among  vertices,
\begin{equation*}
l=(-2,-4;1,1,1,1,1,1),
\end{equation*}
and vertex coordinates of $\Delta^*$ give rise to the GKZ-system  by equation \ref{eq:2.2} ,
\begin{equation}\label{eq:3.1}
\begin{gathered}
\mathcal{Z}_{0,1}= \vartheta_{1,0}+\vartheta_1+\vartheta_2,\quad
\mathcal{Z}_{0,2}=\vartheta_{2,0}+\sum^{6}_{i=3}\vartheta_i,\quad
\mathcal{Z}_i=-\vartheta_i+\vartheta_6,\quad i=1,\ldots,5,\\
\mathcal{L}_1=\prod^6_{i=1}\frac{\partial}{\partial a_i}-(\frac{\partial}{\partial a_{0,1}})^2(\frac{\partial}{\partial a_{0,2}})^4,
\end{gathered}
\end{equation}
with $\vartheta_i=a_i \frac{\partial}{\partial a_i}$ logarithmic derivative. Here $\mathcal{Z}_{0,1}$ and $\mathcal{Z}_{0,2}$  correspond to the invariance of $P_1,P_2$ under overall rescaling ,  $\mathcal{Z}_{i}$  relate to the torus symmetry,
\begin{equation*}
\mathcal{Z}_i:\quad X_i \mapsto \lambda X_i,\quad (a_i,a_6)\mapsto(\lambda^{-1} a_i,\lambda a_6),\quad i=1,\ldots,5,
\end{equation*} 
 and $\mathcal{L}_i$  represents the relations among monomials in $P_1$ and $P_2$, 
\begin{equation*}
\mathcal{L}_1 :\prod_{i=1}^6\frac{\partial}{\partial a_i}(P_1P_2)=(\frac{\partial}{\partial a_{0,1}})^2(\frac{\partial}{\partial a_{0,2}})^4(P_1P_2),
\end{equation*}

The  B-branes wrap on the curve $\Sigma$ on $M_{2,4}$, 
\begin{equation*}
\begin{aligned}
\{P_1=P_2=0\}\cap
\{h_1 = a_7 X_3+ a_8 X_4=0\} \cap\{h_2=  a_9X_3+a_{10}X_5=0\},
\end{aligned}
\end{equation*}
with two extra linear relation $l^1=(0,0;0,0,-1,1,0,0), l^2=(0,0;0,0,-1,0,1,0)$ among vertices.
After blowing up $M_{2,4}$ along the curve $\Sigma$, we obtain the blow-up manifold $M^\prime$ as the complete intersection the projective bundle,
\begin{equation}\label{eq:3.2}
M^\prime:P_1=P_2=0,\quad Q=y_1(a_9X_3+a_{10}X_5)-y_2(a_7X_3+a_8X_4)=0,
\end{equation}
with $(y_1,y_2)$ are homogeneous coordinates on $\mathbb{P}^1$.

The GKZ-system that annihilates period integral of the blow-up $M^\prime$  is obtained from the GKZ-system \ref{eq:3.1} of $M_{2,4}$ by observing the invariance under overall rescalling and torus symmetry of  equation \ref{eq:3.2}. Due to the presence of equation $Q=0$, there are new invariance under overall scalling of $Q=0$, unchanged torus symmetry $\mathcal{Z}_i, i=1,2$ and new torus symmetry,  
\begin{equation*}
\begin{aligned}
\mathcal{Z}_3^\prime:&\quad X_3\mapsto \lambda X_3,\quad(a_3,a_6,a_7,a_9)\mapsto(\lambda^{-1} a_3,\lambda a_6,\lambda^{-1} a_7,\lambda^{-1} a_9)\\[10pt]
 \mathcal{Z}_4^\prime:&\quad X_4\mapsto \lambda X_4,\quad(a_4,a_6,a_8)\mapsto (\lambda^{-1}a_4,\lambda a_6,\lambda^{-1} a_{8}) \\[10pt]
\mathcal{Z}_5^\prime:&\quad X_5\mapsto \lambda X_5,\quad(a_5,a_6,a_{10})\mapsto (\lambda^{-1}a_5,\lambda a_6,\lambda^{-1}a_{10}),\\[10pt]
\mathcal{Z}_6^\prime:&\quad (y_1,y_2)\mapsto(\lambda y_1,\lambda^{-1} y_2).\\
 \end{aligned}
\end{equation*}
 In addition, there are new  algebraic relation between monomials in $Q$,
\begin{equation*}
\begin{gathered}
\frac{\partial (P_1P_2)}{\partial a_4}\frac{\partial Q}{\partial  a_7}=\frac{\partial (P_1P_2)}{\partial a_3}\frac{\partial Q}{\partial a_8}, \\[10pt]
\frac{\partial (P_1P_2)}{\partial a_5}\frac{\partial Q}{\partial  a_9}=\frac{\partial (P_1P_2)}{\partial a_3}\frac{\partial Q}{\partial a_{10}}
\end{gathered}
\end{equation*}
Then, the whole GKZ-system of $M^\prime$ is , 
\begin{equation}
\begin{gathered}
\mathcal{Z}_{0,1}^\prime=\vartheta_{0,1}+\vartheta_1+\vartheta_2,\quad \mathcal{Z}_{0,2}^\prime=\vartheta_{0,2}+\sum^6_{i=3}\vartheta_i,\quad\mathcal{Z}_{3,0}=\sum_{i=7}^{10}\vartheta_i,\\[10pt]
\mathcal{Z}_1=-\vartheta_1+\vartheta_6,\quad\mathcal{Z}_2^\prime=-\vartheta_2+\vartheta_6,\quad\mathcal{Z}_3^\prime=-\vartheta_3+\vartheta_6-\vartheta_7+\vartheta_9,\\[10pt]
 \mathcal{Z}_4^\prime=-\vartheta_4+\vartheta_6-\vartheta_8,
\quad\mathcal{Z}_5^\prime=-\vartheta_5+\vartheta_6-\vartheta_{10},\quad \mathcal{Z}_6^\prime=-\vartheta_7-\vartheta_8+\vartheta_9+\vartheta_{10}\\
\mathcal{L}_1^\prime=\prod^6_{i=1}\frac{\partial}{\partial a_i}-(\frac{\partial}{\partial a_{0,1}})^2(\frac{\partial}{\partial a_{0,2}})^4, \\[10pt]
\mathcal{L}_2^\prime=\frac{\partial}{\partial a_4} \frac{\partial}{\partial a_7}-\frac{\partial}{\partial a_3} \frac{\partial}{\partial a_8},\quad \mathcal{L}_3^\prime=\frac{\partial}{\partial a_5} \frac{\partial}{\partial a_9}-\frac{\partial}{\partial a_3} \frac{\partial}{\partial a_{10}},
\end{gathered}
\end{equation}
which relates to the following  extended polyhedron $\Delta^\prime$,
\begin{table}[H]
\centering
 \begin{tabular}{l|ccccccccc|ccc}
 $\quad$& \multicolumn{9}{|c|}{$\Delta^\prime$}&$l^\prime_1$&$l^\prime_2$&$l^\prime_3$\\
 \hline
 $\nu^\prime_{0,1}$&$1$&$0$&$0$&$0$&$0$&$0$&$0$&$0$&$0$&$-2$&$0$&$0$\\
 $\nu^\prime_{0,2}$&$0$&$1$&$0$&$0$&$0$&$0$&$0$&$0$&$0$&$-4$&$0$&$0$\\
 \color{blue}{$\nu^\prime_{1}$}&$1$&$0$&$0$&$1$&$0$&$0$&$0$&$0$&$0$&$1$&$0$&$0$\\
 \color{blue}{$\nu^\prime_{2}$}&$1$&$0$&$0$&$0$&$1$&$0$&$0$&$0$&$0$&$1$&$0$&$0$\\
\color{blue}{$\nu^\prime_{3}$}&$0$&$1$&$0$&$0$&$0$&$1$&$0$&$0$&$0$&$3$&$-1$&$-1$\\
 \color{blue}{$\nu^\prime_{4}$}&$0$&$1$&$0$&$0$&$0$&$0$&$1$&$0$&$0$&$0$&$1$&$0$\\
\color{blue}{$\nu^\prime_{5}$}&$0$&$1$&$0$&$0$&$0$&$0$&$0$&$1$&$0$&$0$&$0$&$1$\\
 \color{blue}{$\nu^\prime_{6}$}&$0$&$1$&$0$&$-1$&$-1$&$-1$&$-1$&$-1$&$0$&$1$&$0$&$0$\\

 \color{red}{$\nu^\prime_{7}$}&$0$&$0$&$1$&$0$&$0$&$1$&$0$&$0$&$-1$&$-1$&$1$&$0$\\
\color{red}{$\nu^\prime_{8}$}&$0$&$0$&$1$&$0$&$0$&$0$&$1$&$0$&$-1$&$1$&$-1$&$0$\\
\color{red}{$\nu^\prime_{9}$}&$0$&$0$&$1$&$0$&$0$&$1$&$0$&$0$&$1$&$-1$&$0$&$1$\\
\color{red}{$\nu^\prime_{10}$}&$0$&$0$&$1$&$0$&$0$&$0$&$0$&$1$&$1$&$1$&$0$&$-1$\\
 \end{tabular}
 \caption{Toric Information of the Extended Polyhedron}
  \begin{minipage}{\textwidth}
Here $v^\prime_i$ are the integral vertices of the polyhedron, $l^\prime_i$ correspond to the maximal triangulation of $\Delta^\prime$, such that $l=l^\prime_1+ l^\prime_2+l^\prime_3$. The blue vertices denote the original vertices on the polyhedron $\Delta^*$ of $M^*$. The red vertices denote the added vertices for the brane. 
    \end{minipage}
\end{table}

The coordinates $z_i$ on the complex structure moduli space of $M^\prime$, are given  by equation \ref{eq:2.3},
\begin{equation}
z_1=\frac{a_1 a_2 a_3^3  a_6 a_8 a_{10}}{a_{0,1}^2 a_{0,2}^4  a_7 a_9},\quad z_2=\frac{a_4 a_7}{a_3 a_8},\quad z_3=\frac{a_5 a_9}{a_3 a_{10}}
\end{equation}
In terms of  the logarithmic derivatives $\theta_i=z_i\frac{d}{d z_i}$, the operators $\mathcal{L}^\prime_i$ can be rewritten as,  
\begin{equation}\label{eq:3.5}
\begin{aligned}
\mathcal{D}_1&=\theta_1^3 \prod_{k=0}^{2}(3\theta_1-\theta_2-\theta_3-i)(\theta_1-\theta_2)(\theta_1-\theta_3)\\
&-z_1
 (-\theta_1+\theta_2)(-\theta_1+\theta_3)\prod_{i=1}^2 \prod_{j=1}^4(-2\theta_1-i) (-4\theta_1-j),\\[10pt]
 \mathcal{D}_2&= \theta_2(-\theta_1+\theta_2)-z_2
 (\theta_1-\theta_2)  (3\theta_1-\theta_2-\theta_3)\\[10pt]
 \mathcal{D}_3&=\theta_3(-\theta_1+\theta_3)-z_3 (\theta_1-\theta_3) (3\theta_1 -\theta_2-\theta_3),\\
 \cdots
 \end{aligned}
 \end{equation}
where each operator $\mathcal{D}_i$ corresponds to a linear combination among the charge vectors $l^\prime_1,l^\prime_2,l^\prime_3$.

\subsubsection{Brane Superpotential and Disk Instantons} 
 By equation \ref{eq:2.6} ,  we solve the equation \ref{eq:3.5} at $z_i\rightarrow 0$ . The unique series solution, as well as the fundamental period of $M_{2,4}$, is,
\begin{equation*}
\omega_0=1 + 48 z + 15120 z^2 + 7392000 z^3 + \mathcal{O}(z^3),
\end{equation*}
with $z=z_1z_2z_3$. The  solutions  with single-logarithmic leading term, 
\begin{equation*}
\begin{aligned}
\omega_{1,1}=&\omega_0\log(z_1)-24 z_1 z_2 + 1260 z_1^2 z_2^2 - 24 z_1 z_3 + 256 z_1 z_2 z_3 + 
 24 z_1 z_2^2 z_3 + 1260 z_1^2 z_3^2\\& + 24 z_1 z_2 z_3^2
 +\mathcal{O}(z^4)\\[10pt]
\omega_{1,2}=&\omega_0\log(z_2)+24 z_1 z_3 - 24 z_1 z_2^2 z_3 - 1260 z_1^2 z_3^2+\mathcal{O}(z^4)\\[10pt]
\omega_{1,3}=&\omega_0\log(z_3)+24 z_1 z_2 - 1260 z_1^2 z_2^2 - 24 z_1 z_2 z_3^2+\mathcal{O}(z^4)
\end{aligned}
\end{equation*}
give rise to the flat coordinates on the open-closed moduli space by equation \ref{eq:2.5}, 
and mirror map to A-model moduli space,
\begin{equation*}
\begin{aligned}
z_1=&q_1 + 24 q_1^2 q_2 + 24 q_1^2 q_3 - 256 q_1^2 q_2 q_3+\mathcal{O}(q^4)\\[10pt]
z_2=&q_2 - 24 q_1 q_2 q_3 + 24 q_1 q_2^3 q_3 + 972 q_1^2 q_2 q_3^2 +\mathcal{O}(q^5)\\[10pt]
z_3=&q_3 - 24 q_1 q_2 q_3 + 972 q_1^2 q_2^2 q_3 +\mathcal{O}(q^5).
\end{aligned}
\end{equation*}

Furthermore, there are also double logarithmic solutions with leading terms
\begin{equation*}
\begin{gathered}
\frac{1}{2} \ell_1^2,\quad\frac{1}{2}\ell_2^2+\ell_1\ell_2,\quad\frac{1}{2}\ell_3^2+\ell_1\ell_3,\quad\ell_2\ell_3
\end{gathered}
\end{equation*}
where $\log(z_i)$'s are abbreviated as $\ell_i$'s. The B-brane superpotential is given by the linear combination of these solutions , 
\begin{equation}
\mathcal{W}=2(t-t_1)^2+\sum_{N}N_{k,m,n}\mathrm{Li}_2(q_1^{k}q_2^{m}q_3^{n}),
\end{equation}
where $t=t_1+t_2+t_3$ is the flat coordinate of closed B-model and $N_{l,m,n}$ are the Ooguri-Vafa invariants. First a few invariants  are summarized in table \ref{tab:ov24}. They are unchanged under exchange of $m$ and $n$, and the table \ref{tab:ov24} is symmetric along the diagonal, which results from the symmetry of the curve $\Sigma$ that we choose. 

 \subsection{\texorpdfstring{Branes on CICY $M_{3,4}$}{M(3,4)} }
  The next calculation example is studied on the mirror CICY $M_{3,4}\subset\mathbb{P}^5_{(1,1,1,1,1,2)}$. Superpotentials and invariants are obtained.
\subsubsection{GKZ-system and Blowing Up Geometry}
The vertices on the polyhedron $\Delta^*$ associated to $M^*$ are as follow,
\begin{equation*}
\begin{gathered}
\quad\nu^*_1=(1,0,0,0,0),\quad \nu^*_2=(0,1,0,0,0),
\quad \nu^*_3=(0,0,1,0,0),\\[10pt]
\nu^*_4=(0,0,0,1,0),\quad\nu^*_5=(0,0,0,0,1),\quad \nu^*_6=(-1,-1,-1,-1,-2),
\end{gathered}
\end{equation*}
and they satisfy the linear relation,
\begin{equation*}
l=(-3,-4;1,1,1,1,2,1),
\end{equation*}

By the nef-partition, $E_1=\{\nu^*_5,\nu^*_6\}$ and $\{\nu^*_1,\nu^*_2,\nu^*_3,\nu^*_4\}$,  the mirror threefold $M_{3,4}$ is defined by,
\begin{equation*}
\begin{cases}
P_1=a_{0,1}-a_5 X_5-a_6(X_1X_2X_3X_4X_5^2)^{-1}\\[10pt]
P_2=a_{0,2}-a_1 X_1-a_2 X_2- a_3 X_3-a_4X_4, 
\end{cases}
\end{equation*}

The period integral is annihilated by the  GKZ-system,
\begin{equation}\label{eq:3.7}
\begin{gathered}
\mathcal{Z}_{0,1}= \vartheta_{0,1}+\vartheta_5+\vartheta_6,\quad
\mathcal{Z}_{0,2}=\vartheta_{0,2}+\sum^{4}_{i=1}\vartheta_i,\\[10pt]
\mathcal{Z}_i=-\vartheta_i+\vartheta_6,\quad i=1,2,3,4,\quad
\mathcal{Z}_5=-\vartheta_5+2\vartheta_6,\\[10pt]
\mathcal{L}_1=\prod^4_{i=1}\frac{\partial}{\partial a_i}(\frac{\partial}{\partial a_5})^3(\frac{\partial}{\partial a_6})-(\frac{\partial}{\partial a_{1,0}})^3(\frac{\partial}{\partial a_{2,0}})^4,
\end{gathered}
\end{equation}
where $\mathcal{Z}$ are related to the invariance of equation $P_1=P_2=0$ under overall rescaling and torus symmetry, and $\mathcal{L}_i$  corresponds to the relations among Laurent monomials in $P_1$ and $P_2$.
 
Consider the following curve $\Sigma$ on $M_{3,4}$, 
\begin{equation*}
\{P_1=P_2=0\} \cap \{h_1 = a_7 X_1+ a_8 X_2=0\} \cap\{ h_2=  a_9X_1+a_{10}X_3=0\},
\end{equation*}
described by two extra linear relation $l^1=(0,0;-1,1,0,0,0,0),l^2=(0,0;-1,0,1,0,0,0)$ among vertices.

The blow-up manifold $M^\prime$ from blowing up $M_{3,4}$ along the curve $\Sigma$ is defined by,
\begin{equation*}
M^\prime:P_1=P_2=0,\quad Q=y_1h_2-y_2h_1=0.
\end{equation*}
The GKZ-system of $M\prime$ is derived from the GKZ-system of $M_{3,4}$ \ref{eq:3.7},
\begin{equation}
\begin{gathered}
\mathcal{Z}_{0,1}^\prime=\mathcal{Z}_{0,1},\quad \mathcal{Z}_{0,2}^\prime=\mathcal{Z}_{0,2},\quad\mathcal{Z}_{3,0}^\prime=\sum_{i=7}^{10}\vartheta_i,\\[10pt]
\mathcal{Z}_1^\prime=\mathcal{Z}_1-\vartheta_7-\vartheta_9,\quad\mathcal{Z}_2^\prime=\mathcal{Z}_2-\vartheta_8,
\quad \mathcal{Z}_3^\prime=\mathcal{Z}_3-\vartheta_{10},\\[10pt]
\mathcal{Z}_4^\prime=\mathcal{Z}_4,\quad \mathcal{Z}_5^\prime=\mathcal{Z}_4,\quad\mathcal{Z}_6^\prime=-\vartheta_7-\vartheta_8+\vartheta_9+\vartheta_{10}\\[10pt]
\mathcal{L}_1^\prime=\mathcal{L}_1, \\[10pt]
\mathcal{L}_2^\prime=\frac{\partial}{\partial a_2} \frac{\partial}{\partial a_7}-\frac{\partial}{\partial a_{1}} \frac{\partial}{\partial a_8},\quad \mathcal{L}_3^\prime=\frac{\partial}{\partial a_3} \frac{\partial}{\partial a_9}-\frac{\partial}{\partial a_1} \frac{\partial}{\partial a_{10}},
\end{gathered}
\end{equation}
 with the  corresponding  enhanced polyhedron $\Delta^\prime$ given by,
\begin{table}[H]
\centering
 \begin{tabular}{l|ccccccccc|ccc}
 $\quad$& \multicolumn{9}{|c|}{$\Delta^\prime$}&$l^\prime_1$&$l^\prime_2$&$l^\prime_3$\\
 \hline
 $\nu^\prime_{0,1}$&$1$&$0$&$0$&$0$&$0$&$0$&$0$&$0$&$0$&$-3$&$0$&$0$\\
 $\nu^\prime_{0,2}$&$0$&$1$&$0$&$0$&$0$&$0$&$0$&$0$&$0$&$-4$&$0$&$0$\\
 \color{blue}{$\nu^\prime_{1}$}&$0$&$1$&$0$&$1$&$0$&$0$&$0$&$0$&$0$&$3$&$-1$&$-1$\\
  \color{blue}{$\nu^\prime_{2}$}&$0$&$1$&$0$&$0$&$1$&$0$&$0$&$0$&$0$&$0$&$1$&$0$\\
 \color{blue}{$\nu^\prime_{3}$}&$0$&$1$&$0$&$0$&$0$&$1$&$0$&$0$&$0$&$0$&$0$&$1$\\
 \color{blue}{$\nu^\prime_{4}$}&$0$&$1$&$0$&$0$&$0$&$0$&$1$&$0$&$0$&$1$&$0$&$0$\\
  \color{blue}{$\nu^\prime_{5}$}&$1$&$0$&$0$&$0$&$0$&$0$&$0$&$1$&$0$&$2$&$0$&$0$\\
  \color{blue}{$\nu^\prime_{6}$}&$1$&$0$&$0$&$-1$&$-1$&$-1$&$-1$&$-2$&$0$&$1$&$0$&$0$\\
\color{red}{$\nu^\prime_{7}$}&$0$&$0$&$1$&$1$&$0$&$0$&$0$&$0$&$-1$&$-1$&$1$&$0$\\
 \color{red}{$\nu^\prime_{8}$}&$0$&$0$&$1$&$0$&$1$&$0$&$0$&$0$&$-1$&$1$&$-1$&$0$\\
 \color{red}{$\nu^\prime_{9}$}&$0$&$0$&$1$&$1$&$0$&$0$&$0$&$0$&$1$&$-1$&$0$&$1$\\
\color{red}{$\nu^\prime_{10}$}&$0$&$0$&$1$&$0$&$0$&$1$&$0$&$0$&$1$&$1$&$0$&$-1$\\
 \end{tabular}
 \caption{Toric Information of the Extended Polyhedron}
  \begin{minipage}{\textwidth}
Here $v^\prime_i$ are the integral vertices of the polyhedron,$l^\prime_i$correspond to the maximal triangulation of $\Delta^*$,  s.t. $l=l^\prime_1+ l^\prime_2+l^\prime_3$. The blue and red vertices denote the original and added vertices, respectively. 
    \end{minipage}
\end{table}

The coordinates $z_i$ on the complex structure moduli space of $M^\prime$ is obtained by \ref{eq:2.3},
\begin{equation}
z_1=\frac{a_1^3  a_4a_5^2 a_6 a_8 a_{10}}{a_{1,0}^3 a_{2,0}^4 a_2 a_7 a_9},\quad z_2=\frac{a_2 a_7}{a_1 a_8},\quad z_3=\frac{a_3 a_9}{a_1 a_{10}},
\end{equation}

 The corresponding Picard-Fuchs operators $\mathcal{D}_i$ can be obtained by  $\mathcal{L}_i$ operators 
\begin{equation}\label{eq:3.11}
\begin{aligned}
\mathcal{D}_1&=\prod^2_{k=0} (3\theta_1-\theta_2-\theta_3-k)(2\theta^3)(2\theta-1)(\theta_1-\theta_2)(\theta_1-\theta_3)\\
& -z_1
 (-\theta_1+\theta_2)(-\theta_1+\theta_3)\prod_{i=1}^3 \prod_{j=1}^4(-3\theta_1-i) (-4\theta_1-j),\\[10pt]
 \mathcal{D}_2&= \theta_2 (-\theta_1+\theta_2)-z_2
 (3\theta_1-\theta_2-\theta_3)  (\theta_1-\theta_2)\\[10pt]
 \mathcal{D}_3&=\theta_3(-\theta_1+\theta_3)-z_3 (3\theta_1-\theta_2-\theta_3) (\theta_1 -\theta_3),\\
 \cdots
 \end{aligned}
 \end{equation}

\subsubsection{Brane Superpotential and Disk Instantons}  
 The solutions of equation \ref{eq:3.11} underlie the mirror map and the superpotential.
 The series solution
\begin{equation*}
\omega_0=1 + 72 z + 37800 z^2 + 31046400 z^3 + \mathcal{O}(z^{3}),
\end{equation*}
with $z=z_1z_2z_3$, together with the following single logarithmic solutions,
\begin{equation*}
\begin{aligned}
\omega_{1,1}=&\omega_0\log(z_1)-36 z_1 z_2 + 3150 z_1^2 z_2^2 - 36 z_1 z_3 + 420 z_1 z_2 z_3 + 
 36 z_1 z_2^2 z_3\\& + 3150 z_1^2 z_3^2 + 36 z_1 z_2 z_3^2+\mathcal{O}(z^4)\\[10pt]
\omega_{1,2}=&\omega_0\log(z_2)+36 z_1 z_3 - 36 z_1 z_2^2 z_3 - 3150 z_1^2 z_3^2  +\mathcal{O}(z^4)\\[10pt]
\omega_{1,3}=&\omega_0\log(z_3)+ 36 z_1 z_2 - 3150 z_1^2 z_2^2 - 36 z_1 z_2 z_3^2+\mathcal{O}(z^4)
\end{aligned}
\end{equation*}
give rise to the mirror map, connceting the B-model complex moduli $z$ to the A-model moduli $q$,
\begin{equation*}
\begin{aligned}
z_1=&q_1 + 36 q_1^2 q_2 - 1206 q_1^3 q_2^2 + 36 q_1^2 q_3 - 420 q_1^2 q_2 q_3 + 
 1296 q_1^3 q_2 q_3 - 36 q_1^2 q_2^2 q_3 \\&- 1206 q_1^3 q_3^2 - 36 q_1^2 q_2 q_3^2+\mathcal{O}(q^5)\\[10pt]
z_2=&q_2 - 36 q_1 q_2 q_3 + 36 q_1 q_2^3 q_3 +\mathcal{O}(q^5)\\[10pt]
z_3=&q_3 - 36 q_1 q_2 q_3 + 2502 q_1^2 q_2^2 q_3 + 36 q_1 q_2 q_3^3+\mathcal{O}(q^5)
\end{aligned}
\end{equation*}
In addition, the superpotential is given by the linear combinations of double logarithmic solutions, 
\begin{equation}
\mathcal{W}=2(t-t_1)^2+\sum_{N}N_{k,m,n}\mathrm{Li}_2(q_1^{k}q_2^{m}q_3^{n}),\\
\end{equation}
 First a few invariants are summarized in table  \ref{tab:ov34}. 
 
 \subsection{\texorpdfstring{Branes on CICY $M_{2,6}$, $M_{3,6}$, and $M_{4,6}$}{M(2,6)M(3,6)M(4,6)} }
 The B-brane superpotentials on CICY on $M_{2,6}\subset \mathbb{P}^5_{(1,1,1,1,3)}$,$ M_{3,6}\subset \mathbb{P}^5_{(1,1,1,2,3)}$, and $M_{2,6}\subset \mathbb{P}^5_{(1,1,2,2,3)}$ are computed and the Ooguri-Vafa invariants on the corresponding A-model threefold $M^*$ are extracted. For brevity,  certain technical details that are similar to the last two examples are omitted. The GKZ-system of blow-ups are summarized in Appendix  \ref{gkz:blp} and invariants in Appendix \ref{ov}.

\section{Summary and Conclusions}
In this work, we calculate the B-brane superpotententials and Ooguri-Vafa invariants with several deformations on compact complete intersection Calabi-Yau threefolds by blowing up the curve wrapped by B-branes. The complex structure moduli  of the blow-up manifolds relates  complex structure moduli and brane moduli on the original threefolds $M$. From  observation on the defining equation of $M^\prime$, the GKZ-system of blow-up is obtained from the GKZ-system associated to $M$, which  annihilates the period matrix of $M^\prime$.  The  logarithmic solutions are used to construct  mirror maps and B-brane superpotentials.  By multi-cover formula and  mirror symmetry, the Ooguri-Vafa invariants are extracted from A-model side and interpreted as counting disk instantons. In our calculation,  suitable A-brane configurations on A-model $M$ are assumed to exist.  An independent computation directly in the topological A-models would further support our results by mirror symmetry.

The one-parameter CICY  threefolds in this paper can be obtained by extremal transition from multi-parameter models\cite{Katz1996,Klemm1996,Morrison1997}. It would be interesting to calculate the Ooguri-Vafa invariants on more complicated models and study the relation of the invariants on two distinct models. 
In addition,  the arithmetic properties of  the mirror maps and the superpotentials are worth to study.

\section*{Acknowledgement}
  This work is dedicated to our dear supervisor Prof. Fu-Zhong Yang 
 who  sadly passed away while the paper was being prepared.

\appendix

\section{GKZ-system of Blow-ups}\label{gkz:blp}
\subsection{\texorpdfstring{CICY $M_{2,6}$}{M(2,6)} }
The brane geometry is captured by the following linear relation,
 
\begin{equation*}
  \begin{gathered}
 l=(-2,-6;1,1,1,1,3,1), \\[10pt]
 l^1=(0,0;0,1,-1,0,0,0),\quad  l^2=(0,0;0,1,0,-1,0,0),
  \end{gathered}  
\end{equation*}
The GKZ-system associated to the blow-up $M^\prime$ is,
\begin{equation}\label{eq:A.1}
\begin{gathered}
\mathcal{Z}_{1,0}^\prime=\vartheta_{1,0}+\vartheta_1+\vartheta_6,\quad \mathcal{Z}_{2,0}^\prime=\vartheta_{2,0}+\sum^5_{i=2}\vartheta_i,\quad\mathcal{Z}_{3,0}=\sum_{i=7}^{10}\vartheta_i,\\[10pt]
\mathcal{Z}_1=-\vartheta_1+\vartheta_6,\quad\mathcal{Z}_2^\prime=-\vartheta_2+\vartheta_6+\vartheta_7+\vartheta_9,\quad\mathcal{Z}_3^\prime=-\vartheta_3+\vartheta_6+\vartheta_8,\\[10pt]
 \mathcal{Z}_4^\prime=-\vartheta_4+\vartheta_6+\vartheta_{10},
\quad\mathcal{Z}_5^\prime=-\vartheta_5+3\vartheta_6,\quad \mathcal{Z}_6^\prime=-\vartheta_7-\vartheta_8+\vartheta_9+\vartheta_{10},\\[10pt]
\mathcal{L}_1^\prime=\prod^4_{i=1}\frac{\partial}{\partial a_i}(\frac{\partial}{\partial a_5})^3(\frac{\partial}{\partial a_6})-(\frac{\partial}{\partial a_{0,1}})^2(\frac{\partial}{\partial a_{0,2}})^6, \\[10pt]
\mathcal{L}_2^\prime=\frac{\partial}{\partial a_3} \frac{\partial}{\partial a_7}-\frac{\partial}{\partial a_2} \frac{\partial}{\partial a_8},\quad \mathcal{L}_3^\prime=\frac{\partial}{\partial a_4} \frac{\partial}{\partial a_9}-\frac{\partial}{\partial a_2} \frac{\partial}{\partial a_{10}},
\end{gathered}
\end{equation}
with the relation among vertices of the enhanced polyhedron $\Delta^\prime$,
\begin{equation*}
  \begin{gathered}
 l_1^\prime=(-2,-6;1,3,0,0,3,1,-1,1-1,1), \\[10pt]
 l^\prime_2=(0,0;0,-1,1,0,0,0,1,-1,0,0),\quad  l^\prime_3=(0,0;0,-1,0,1,0,0,0,0,1,-1).
  \end{gathered}  
\end{equation*}

The B-brane superpotential is obtained by solve the GKZ-system \ref{eq:A.1} and Ooguri-Vafa invariants are extracted at the lorge volume point of A-model as in table \ref{tab:ov26}.

\subsection{\texorpdfstring{CICY $M_{3,6}$}{M(3,6)} }
The brane geometry is described by,
\begin{equation*}
    \begin{gathered}
    l=(-3,-6;1,1,1,2,3,1),\\[10pt]
    l^1=(0,0;1,-1,0,0,0,0),\quad l^2=(0,0;1,0,-1,0,0,0)
    \end{gathered}
\end{equation*}

 The GKZ-system of the blow-up $M^\prime$,
\begin{equation}\label{eq:A.2}
\begin{gathered}
\mathcal{Z}_{0,1}^\prime=\vartheta_{1,0}+\sum_{i=1}^3\vartheta_i,\quad \mathcal{Z}_{0,2}^\prime=\vartheta_{2,0}+\sum^6_{i=4}\vartheta_i,\quad\mathcal{Z}_{3,0}^\prime=\sum_{i=7}^{10}\vartheta_i,\\[10pt]
\mathcal{Z}_1^\prime=-\vartheta_1+\vartheta_6-\vartheta_7-\vartheta_9,\quad\mathcal{Z}_2^\prime=-\vartheta_2+\vartheta_6-\vartheta_{8},
\quad \mathcal{Z}_3^\prime=-\vartheta_3+\vartheta_6-\vartheta_{10},\\[10pt]
\mathcal{Z}_4^\prime=-\vartheta_4+2\vartheta_6,\quad \mathcal{Z}_5^\prime=-\vartheta_5+3\vartheta_6,\quad\mathcal{Z}_6^\prime=-\vartheta_7-\vartheta_8+\vartheta_9+\vartheta_{10}\\[10pt]
\mathcal{L}_1^\prime=\prod^{3}_{i=1}(\frac{\partial}{\partial a_i})(\frac{\partial }{\partial a_4})^2(\frac{\partial}{\partial a_5})^3(\frac{\partial}{\partial a_6})-(\frac{\partial}{\partial a_{0,1}})^3(\frac{\partial}{\partial a_{0,2}})^6, \\[10pt]
\mathcal{L}_2^\prime=\frac{\partial}{\partial a_2} \frac{\partial}{\partial a_7}-\frac{\partial}{\partial a_1} \frac{\partial}{\partial a_8},\quad \mathcal{L}_3^\prime=\frac{\partial}{\partial a_3} \frac{\partial}{\partial a_9}-\frac{\partial}{\partial a_1} \frac{\partial}{\partial a_{10}},
\end{gathered}
\end{equation}
with the maximal triangulation of thee enhanced polyhedron corresponding to,
\begin{equation*}
    \begin{gathered}
    l=(-3,-6;3,0,0,2,3,1,-1,1,-1,1),\\[10pt]
    l^1=(0,0;-1,1,0,0,0,0,1,-1,0,0),\quad l^2=(0,0;-1,0,1,0,0,0,0,0,1,-1)
    \end{gathered}
\end{equation*}
The superpotential and invariantes are given by solving the above equation \ref{eq:A.2}. The results are summarizeed in the table \ref{tab:ov36}.

\subsection{\texorpdfstring{CICY $M_{4,6}$}{M(4,6)} }
The brane geometry is give by specifing two extra linear relation,
\begin{equation*}
\begin{gathered}
l=(-4,-6;1,1,2,2,3,1),\\[10pt]
l^1=(0,0;1,-1,0,0,0,0),\quad l^2=(0,0;1,0,0,0,0,-1),
\end{gathered}
\end{equation*}
 The GKZ-system associated to the blow-up $M^\prime$ is,
\begin{equation}
\begin{gathered}
\mathcal{Z}_{0,1}^\prime=\vartheta_{0,1}+\vartheta_3+\vartheta_4,\quad \mathcal{Z}_{0,2}^\prime=\vartheta_{0,2}+\sum^6_{i=4}\vartheta_i,\quad\mathcal{Z}_{0,3}^\prime=\sum_{i=7}^{10}\vartheta_i,\\[10pt]
\mathcal{Z}_1^\prime=-\vartheta_1+\vartheta_6-\vartheta_7-\vartheta_9-\vartheta_{10},\quad\mathcal{Z}_2^\prime=-\vartheta_2+\vartheta_6-\vartheta_8-\vartheta_{10},
\quad \mathcal{Z}_3^\prime=-\vartheta_3+2\vartheta_6+2\vartheta_{10},\\[10pt]
\mathcal{Z}_4^\prime=-\vartheta_4+2\vartheta_6+2\vartheta_{10},\quad \mathcal{Z}_5^\prime=-\vartheta_5+3\vartheta_6+3\vartheta_{10},\quad\mathcal{Z}_6^\prime=-\vartheta_7-\vartheta_8+\vartheta_9+\vartheta_{10}\\[10pt]
\mathcal{L}_1^\prime=\frac{\partial}{\partial a_1}\frac{\partial}{\partial a_2}(\frac{\partial}{\partial a_3})^2(\frac{\partial}{\partial a_4})^2(\frac{\partial}{\partial a_5})^3\frac{\partial}{\partial a_6}-(\frac{\partial}{\partial a_{0,1}})^4(\frac{\partial}{\partial a_{0,2}})^6, \\[10pt]
\mathcal{L}_2^\prime=\frac{\partial}{\partial a_2} \frac{\partial}{\partial a_7}-\frac{\partial}{\partial a_1} \frac{\partial}{\partial a_8},\quad \mathcal{L}_3^\prime=\frac{\partial}{\partial a_6} \frac{\partial}{\partial a_9}-\frac{\partial}{\partial a_1} \frac{\partial}{\partial a_{10}},
\end{gathered}
\end{equation}
with the maximal triangulation of  the extended polyhedron,
\begin{equation*}
\begin{gathered}
l=(-4,-6;3,0,2,2,3,0,-1,1,-1,1),\\[10pt]
l^1=(0,0;-1,1,0,0,0,0,1,-1,0,0),\quad l^2=(0,0;1,0,0,0,0,-1,0,0,1,-1),
\end{gathered}
\end{equation*}
The corresponding Ooguri-Vafa invariants on $M^*_{4,6}$ are listed in table \ref{tab:ov46}.

\section{Ooguri-Vafa Invariants}\label{ov}
\begin{table}[H]
\centering
 \begin{tabular}{c|ccccc}
$k=1$&$n=0$&$1$&$2$&$3$&$4$\\
 \hline
$m=0$&$32$&$	288$&$	-96$&$	16$&$	0$\\
$1$&$288	$&$192	$&$-144$&$	0$&$	0$\\
$2$&$-96$&$	-144$&$	0$&$	0$&$	0$\\
$3$&$16$&$	0$&$	0$&$	0$&$	0$\\
$4$&$0$&$	0$&$0$&$	0$&$	0$\\
\multicolumn{6}{c}{$\quad$}\\
 $k=2$&$n=0$&$1$&$2$&$3$&$4$\\
 \hline
 $m=0$&$160	$&$-1248$&$	-10272$&$	4416	$&$-2112	$\\
$1$&$-1248$&$	8128$&$	9120$&$	-9600$&$	2912$\\
$2$&$-10272	$&$9120$&$	10272$&$	-15456$&$	2808	$\\
$3$&$4416$&$	-9600$&$	-15456$&$	11136$&$	0$\\
$4$&$-2112$&$	2912$&$	2808$&$	0$&$	0$\\
\multicolumn{6}{c}{$\quad$}\\
 $k=3$&$n=0$&$1$&$2$&$3$&$4$\\
 \hline
$m=0$&$1952$&$	-18336$&$	101760$&$	861024	$&$-378624$\\
$1$&$-18336	$&$144384$&$	-691584$&$	-1089280$&$	1283136$\\
$2$&$101760$&$	-691584	$&$2445312$&$	3711808$&$	-2563968$\\
$3$&$861024$&$	-1089280$&$	3711808$&$	3466560	$&$-4092928		$\\
$4$&$-378624$&$	1283136	$&$-2563968	$&$-4092928	$&$2840448	$\\
\end{tabular}
\caption{Ooguri-Vafa Invariants on $M_{2,4}$}
\label{tab:ov24}
\end{table}

\begin{table}[H]
\centering
 \begin{tabular}{c|ccccc}
$k=1$&$n=0$&$1$&$2$&$3$&$4$\\
 \hline
$m=0$&$48	$&$432$&$	-144	$&$24$&$	0$\\
$1$&$432$&$	288$&$	-216$&$	0$&$	0$\\
$2$&$-144$&$	-216	$&$0	$&$0$&$	0$\\
$3$&$24$&$	0$&$	0$&$	0$&$	0$\\
$4$&$0$&$	0$&$	0$&$	0$&$	0$\\
\multicolumn{6}{c}{$\quad$}\\
 $k=2$&$n=0$&$1$&$2$&$3$&$4$\\
 \hline
$m=0$&$408$&$	-3312$&$	-27048$&$	11616$&$	-5400$\\
$1$&$-3312$&$	21600$&$	29328$&$	-26496$&$	8208$\\
$2$&$-27048	$&$29328	$&$31680	$&$-42960$&$	7656$\\
$3$&$11616	$&$-26496	$&$-42960	$&$28416	$&$0$	\\
$4$&$-5400	$&$8208$&$	7656	$&$0$&$	0$\\
\multicolumn{6}{c}{$\quad$}\\
 $k=3$&$n=0$&$1$&$2$&$3$&$4$\\
 \hline
$m=0$&$8208$&$	-80640$&$	466272$&$	3910320$&$	-1721376$\\
$1$&$-80640$&$	646848$&$	-3157056$&$	-5718912$&$	5931072$	\\
$2$&$466272$&$	-3157056	$&$11239776$&$	17265264	$&$-11883456$\\
$3$&$3910320$&$	-5718912$&$	17265264$&$	16301088	$&$-19212576$\\
$4$&$-1721376	$&$5931072	$&$-11883456$&$	-19212576	$&$13049280$\\
\end{tabular}
\caption{Ooguri-Vafa Invariants on $M_{3,4}$ }
\label{tab:ov34}
\end{table}

\begin{table}[H]
\centering
 \begin{tabular}{c|ccccc}
$k=1$&$n=0$&$1$&$2$&$3$&$4$\\
 \hline
$m=0$&$ 160$&$	1440	$&$-480$&$	80$&$	0$\\
$1$&$1440$&$	960$&$	-720$&$	0$&$	0$\\
$2$&$-480$&$	-720$&$	0$&$	0$&$	0$\\
$3$&$80$&$	0$&$	0$&$	0$&$	0$\\
$4$&$0$&$	0$&$0$&$	0$&$	0$\\
\multicolumn{6}{c}{$\quad$}\\
 $k=2$&$n=0$&$1$&$2$&$3$&$4$\\
 \hline
 $m=0$&$5504	$&$-47328$&$	-383040$&$	164160$&$	-73440	$\\
$1$&$-47328$&$	292800$&$	362400$&$	-347520$&$	117600$\\
$2$&$-383040$&$	362400$&$	432480$&$	-574560$&$	113160	$\\
$3$&$164160$&$	-347520$&$	-574560$&$	385920$&$	0$\\
$4$&$-73440$&$	117600$&$	113160$&$	0$&$	0$\\
\multicolumn{6}{c}{$\quad$}\\
 $k=3$&$n=0$&$1$&$2$&$3$&$4$\\
 \hline
$m=0$&$432160$&$	-4503840$&$	27511680$&$	228161760$&$	-100550400$\\
$1$&$-4503840$&$	35859456	$&$-175001472$&$	-305181440$&$	317586240$\\
$2$&$27511680$&$	-175001472$&$	594892800$&$	907979840$&$	-616498560$\\
$3$&$228161760$&$	-305181440$&$	907979840$&$	866126400$&$	-1014325760$\\
$4$&$-100550400$&$	317586240$&$	-616498560$&$	-1014325760	$&$685572480	$\\
\end{tabular}
\caption{Ooguri-Vafa Invariants on $M_{2,6}$}
\label{tab:ov26}
\end{table}

\begin{table}[H]
\centering
 \begin{tabular}{c|ccccc}
$k=1$&$n=0$&$1$&$2$&$3$&$4$\\
 \hline
$m=0$&$240$&$	2160$&$	-720	$&$120	$&$0$\\
$1$&$2160	$&$1440$&$	-1080$&$	0	$&$0$\\
$2$&$-720$&$	-1080$&$	0$&$	0$&$	0$\\
$3$&$120$&$	0$&$	0$&$	0$&$	0$\\
$4$&$0$&$	0$&$	0$&$	0$&$	0$\\
\multicolumn{6}{c}{$\quad$}\\
 $k=2$&$n=0$&$1$&$2$&$3$&$4$\\
 \hline
$m=0$&$13800	$&$-123120$&$	-993240$&$	424800	$&$-186120$\\
$1$&$-123120$&$	781920$&$	1230480	$&$-984960$&$	326160$\\
$2$&$-993240$&$	1230480	$&$1337760$&$	-1625040$&$	299520$\\
$3$&$424800	$&$-984960	$&$-1625040$&$	979200$&$0$	\\
$4$&$-186120	$&$326160$&$	299520	$&$0$&$	0$\\
\multicolumn{6}{c}{$\quad$}\\
 $k=3$&$n=0$&$1$&$2$&$3$&$4$\\
 \hline
$m=0$&$1815120$&$	-19514880$&$	122876640$&$	1013165424	$&$-446689440$\\
$1$&$-19514880$&$	160125120$&$	-801541440$&$	-1655884416	$&$1502625600$	\\
$2$&$122876640	$&$-801541440	$&$2808833760	$&$4414021680	$&$-2977110720$\\
$3$&$1013165424	$&$-1655884416$&$	4414021680$&$	4250616480	$&$-4938103200	$\\
$4$&$-446689440$&$	1502625600	$&$-2977110720$&$	-4938103200	$&$3233476800$\\
\end{tabular}
\caption{Ooguri-Vafa Invariants on $M_{3,6}$}
\label{tab:ov36}
\end{table}

\begin{table}[H]
\centering
  \begin{tabular}{c|ccccc}
$k=1$&$n=0$&$1$&$2$&$3$&$4$\\
 \hline
$m=0$&$480	$&$4320$&$	-1440$&$	240$&$	0$\\
$1$&$4320$&$	2880$&$	-2160$&$	0$&$	0$\\
$2$&$-1440$&$	-2160$&$	0$&$	0$&$	0$\\
$3$&$240$&$	0$&$	0$&$	0$&$	0$\\
$4$&$0$&$	0$&$	0$&$	0$&$	0$\\	
\multicolumn{6}{c}{$\quad$}\\
 $k=2$&$n=0$&$1$&$2$&$3$&$4$\\
 \hline
 $m=1$&$64560$&$	-603360$&$	-4849680$&$	2068800$&$	-883440$\\
$1$&$-603360	$&$3948480	$&$7689120	$&$-5293440$&$	1715040$\\
$2$&$-4849680$&$	7689120$&$	7781760$&$	-8715360$&$	1497840$\\
$3$&$2068800	$&$-5293440$&$	-8715360$&$	4656000$&$	0$\\
$4$&$-883440	$&$1715040$&$	1497840$&$	0$&$	0$\\
\multicolumn{6}{c}{$\quad$}\\
 $k=3$&$n=0$&$1$&$2$&$3$&$4$\\
 \hline
$m=0$&$19966560$&$	-222979680$&$	1458023040$&$	11939377824$&	$-5265987840$\\
$1$&$-222979680$&$	1896606720$&$	-9795738240$&$	-23758008576$&$	19015358400$\\
$2$&$1458023040$&$	-9795738240$&$	35610071040$&$	59297784000	$&$-39000389760$\\
$3$&$11939377824	$&$-23758008576$&$	59297784000$&$	57246494400$&$	-65195381760$\\
$4$&$-5265987840$&$	19015358400$&$	-39000389760$&$	-65195381760	$&$40631529600$\\
\end{tabular}
\caption{Ooguri-Vafa Invariants on $M_{4,6}$}
\label{tab:ov46}
\end{table}

\printbibliography
\end{document}